\documentclass[aps,10pt,prd,twocolumn,preprintnumbers,amsmath,amssymb,nofootinbib,superscriptaddress,a4paper,showpacs]{revtex4-1}

\usepackage[breaklinks]{hyperref}
\usepackage[utf8]{inputenc}
\usepackage{graphicx}
\usepackage{color}
\usepackage{amsfonts,amsthm}
\usepackage{bm,bbm}

\newcommand{\be}{\begin{equation}}
\newcommand{\ee}{\end{equation}}
\newcommand{\ba}{\begin{eqnarray}}
\newcommand{\ea}{\end{eqnarray}}
\def\bs{\begin{subequations}}
\def\es{\end{subequations}}
\def\a{\alpha}
\def\b{\beta}
\def\de{\delta}
\def\g{\gamma}

\def\om{\omega}
\def\De{\Delta}

\newcommand{\Eq}[1]{(\ref{#1})}
\def\com{\color{magenta}}
\def\cob{\color{blue}}
\newcommand{\oarX}[1]{\href{http://arxiv.org/abs/#1}{{\ttfamily\com arXiv:#1}}}
\newcommand{\arX}[1]{\href{http://arxiv.org/abs/#1}{{\ttfamily\com arXiv:#1}}}
\newcommand{\doin}[6]{\href{http://dx.doi.org/#1}{{\cob #2 #3 {\bf #4}, #5 (#6)}}}
\newcommand{\doinn}[5]{\href{http://dx.doi.org/#1}{{\cob #2 {\bf #3}, #4 (#5)}}}

\newcommand{\tia}[1]{}

\def\mp{M_{\rm Pl}}
\def\rme{e}
\def\rmd{d}

\begin{document}

\title{What gravity waves are telling about quantum spacetime}

\author{Michele Arzano}
\email{michele.arzano@roma1.infn.it}
\affiliation{Dipartimento di Fisica and INFN, ``Sapienza'' University of Rome, P.le A.\ Moro 2, 00185 Roma, Italy}

\author{Gianluca Calcagni}
\email{calcagni@iem.cfmac.csic.es}
\affiliation{Instituto de Estructura de la Materia, CSIC, Serrano 121, 28006 Madrid, Spain}

\date{April 1, 2016}

\begin{abstract}
We discuss various modified dispersion relations  motivated by quantum gravity which might affect the propagation of the recently observed gravitational-wave signal of the event GW150914. We find that the bounds set by the data on the characteristic quantum-gravity mass scale $M$ are too weak to constrain these scenarios and, in general, much weaker than the expected $M> 10^4\,\text{eV}$ for a correction to the dispersion relation linear in $1/M$. We illustrate this issue by giving lower bounds on $M$, plus an upper bound coming from constraints on the size of a quantum ergosphere. We also show that a phenomenological dispersion relation $\om^2 = k^2(1+\alpha k^n/M^n)$ is compatible with observations and, at the same time, has a phenomenologically viable mass $M>10\,\text{TeV}$ only in the quite restrictive range $0<n<0.68$. Remarkably, this is the domain of multiscale spacetimes but not of known quantum-gravity models.
\end{abstract}



\pacs{04.70.Dy, 04.60.Bc, 11.10.Lm}

\preprint{\doin{10.1103/PhysRevD.93.124065}{PHYSICAL REVIEW}{D}{93}{124065}{2016} \hspace{9cm} \arX{1604.00541}}

\maketitle

With the discovery of the Higgs boson at the LHC, the notable restriction of the inflationary parameter space by \textsc{Planck}, and the first observation of gravitational waves of astrophysical origin by LIGO, we have entered a veritable gold mine of data that we can use to constrain theories beyond both the Standard Model and general relativity. In particular, the recent constraint on Lorentz violations from the gravitational waves of the event GW150914 \cite{Abb16,LIG41} were used in Ref.\ \cite{EMNan} to bound generic quantum-gravity effects. Given a dispersion relation $\om^2=\om^2(k)$ in $\hbar=1=c$ units, the velocity of propagation of a wave front is given by the group velocity
\be\label{groupv}
v:=\frac{\rmd\om}{\rmd k}\,.
\ee
For the usual Lorentz-invariant massive dispersion relation $\om^2=k^2+m^2$, one gets
\be\label{lorin}
v=\frac{k}{\om}=\sqrt{1-\frac{m^2}{\om^2}}\simeq 1-\frac{m^2}{2\om^2}\,,
\ee
where in the last step we assumed a small mass. The difference $\De v:= v-1\simeq v-k/\om$ between the propagation speed of the signal and the speed of light is approximately equal to
\be\label{devgr}
\De v\simeq -\frac{m^2}{2\om^2}\,.
\ee
The signal of GW150914 is peaked at frequencies (energies) $\om\approx 100\,{\rm Hz}\approx 6.6\times 10^{-14}\,\text{eV}$. In Ref.\ \cite{Abb16,LIG41}, an upper bound for the mass of the graviton was found from Eq.\ \Eq{devgr}, $m<1.2\times 10^{-22}\,{\rm eV}$, leading to
\be\label{devbo}
|\De v| < 1.7 \times 10^{-18}\,.
\ee

From general quantum-gravity arguments, a phenomenological dispersion relation 
\be\label{lindi}
\om^2=k^2-\frac{2b_1}{3}\frac{k^3}{M}
\ee
was obtained in Ref.~\cite{ACEMN} such that the correction to the velocity is linear in the frequency:
\be\label{devqg}
\De v_\textsc{qg} \simeq - b_1 \frac{\om}{M}\,,
\ee
where $b_1=O(1)$ is an unspecified constant factor and $M$ is a mass scale above which new physics arises. Comparing Eqs.\ \Eq{devbo} and \Eq{devqg}, one gets
\be\label{bouM}
M> 4\times 10^4\,{\rm eV}
\ee
(rounded up to $10^5\,{\rm eV}$ in \cite{EMNan}). This constraint is quite weak, in fact much weaker than that coming from photon propagation from gamma-ray bursts \cite{Fermi,HESS,Vas13}. The point in \cite{EMNan} was to illustrate a first example of independent information on Lorentz invariance coming from gravitational waves.

In this paper, however, we show how such information might be virtually negligible in actual examples of quantum-gravity effects near a black hole. As a dimensional argument would easily show, the low frequency of astrophysical gravitational waves is too low to constrain quantum-gravity horizon effects. This is one of the main messages we wish to convey here. We take two specific examples, a modified dispersion relation derived from a nonlocal effective field theory reproducing the black-hole entropy-area law and one motivated by (loop) quantum gravity. Then, we place a bound on deviations from standard physics that could occur in the proximity of a black hole. This bound is 8 orders of magnitude weaker than \Eq{bouM}.

A preliminary remark may help the reader to frame the physical picture in which our, as well as past, results on the subject of modified dispersion relations should be interpreted. Strong gravity effects are only present at the source, during the merger, when gravity waves are produced. What we will effectively model using various forms of dispersion relations are quantum spacetime effects which are present during the propagation of gravitons throughout their whole trajectory, also in flat spacetime. In other words, once a dispersion relation is modified, such modification is ever present in the propagation of a particle and this fact is independent of whether new effects are small or large for the observer detecting the signal. Conceptually, there is no great difference with respect to considering a mass for the graviton \cite{Wil97}, apart from the fact that, in the case of quantum gravity, the ``mass'' correction is energy dependent. This is true not only for gravitational waves, but also for photons emitted by distant sources such as gamma-ray bursts. This point of view is indeed the one commonly adopted in the literature on phenomenology of quantum gravity based on the propagation of photons (see, e.g, Ref.\ \cite{Ame08}) and which has been already applied to the propagation of gravity waves in Ref.\ \cite{EMNan} and earlier works. 

Let us start by considering the $D$-dimensional partition function $Z=\int^{+\infty}_{-\infty}\rmd^Dk\,\rme^{-\b\om(k)}\propto\int_0^{+\infty}\rmd k\,k^{D-1}\,\rme^{-\b\om(k)}$, which can also be written in terms of an energy distribution $\varrho(\om)$, $Z=\int_0^{+\infty}\rmd \om\,\varrho(\om)\,\rme^{-\b\om}$. Equating these two expressions and using the dispersion relation $\om=\om(k)$, one finds that $\varrho(\om)=[k(\om)]^{D-1}/v$, where $v$ is given by Eq.\ \Eq{groupv}. To reproduce the Bekenstein--Hawking entropy-area law of black-hole thermodynamics \cite{BH}, one needs a distribution of (quantum) spacetime microstates that strongly deviates from the one we would obtain in a ordinary local field theory, $\varrho(\om)=\om^3$ in $D=4$. The new distribution $\varrho(\om)$ is associated with a nonlocal field theory characterized by the dispersion relation in four dimensions \cite{Pad98,Pad99}
\be\label{dire}
\om^2=\frac{3M^2}{8\pi}\,\ln\left(1+\frac{8\pi k^2}{3M^2}\right)\,,
\ee
where 
$\om=k^0=E$ is the energy, $k=|{\bf k}|$ is the length of the spatial momentum vector and $M$ (ideally, the reduced Planck mass $\mp$) is the scale at which nonlocal effects become important. In our convention, spacetime has signature $(-,+,+,+)$. From Eq.\ \Eq{dire}, $v=(k/\om)/(1+a^2k^2)$, where $a=\sqrt{8\pi/(3M^2)}$. Plugging this into the expression for the density of states, we obtain $\varrho(\om)=(\om/a^2)(\rme^{a^2\om^2}-1)\rme^{a^2\om^2}\sim \om^3\,\rme^{a^2\om^2}$. Then, from $S\sim\ln\varrho= (a\om)^2+O(\ln\om)$, one finds that the entropy $S$ is proportional to the area of the black-hole horizon plus a logarithmic correction.

The quantum-gravitational degrees of freedom encoded in this effective nonlocal model organize themselves in a way such that the geometry of spacetime is heavily distorted in the ultraviolet \cite{ArCa1}. In the infrared, one has a correction to the massless dispersion relation $\om^2=k^2$, hence a modification to the propagation speed of gravitons. Using Eqs.\ \Eq{groupv} and \Eq{dire} and expanding at small momenta $k\ll M$, we have
\be
v=\frac{k}{\om}\left(1+\frac{8\pi k^2}{3M^2}\right)^{-1}\simeq \frac{k}{\om}\left(1-\frac{8\pi k^2}{3M^2}\right)\,,
\ee
so that
\be\label{devbh}
\De v \simeq -\frac{8\pi}{3}\left(\frac{\om}{M}\right)^2\,.
\ee
Note that the dispersion relation \Eq{dire} we used for the calculation arises from quantum-gravity effects near a black-hole horizon, but it is not based on a black-hole metric background (the only ingredient was to implement the entropy-area law in an effective nonlocal field theory mimicking quantum-gravity physics). Therefore, as in the case of \Eq{devqg}, the effect \Eq{devbh} cannot be naively ascribed to the local curvature of the background on which the wave propagates.

Since the dispersion relation \Eq{dire} only affects gravitational degrees of freedom, we can apply it directly to the regime of linearized gravity which describes the production and propagation of waves. Comparing with Eq.\ \Eq{devbo}, we obtain
\be\label{epin1}
M> \sqrt{\frac{8\pi}{3|\De v|}}\,\om \approx 10^{-4}\,{\rm eV}\,.
\ee
This constraint is much less impressive than the already weak bound \Eq{bouM} obtained in \cite{EMNan} for the linear correction \Eq{devqg}. However, this is what we get in a specific effective model of quantum-spacetime microscopic degrees of freedom. In \cite{Pad98,Pad99}, the relation \eqref{dire} was proposed as one of the possible choices of modified dispersion relations which exhibit the $\ln k^2$ ultraviolet (UV) behavior needed to match the Bekenstein--Hawking entropy-area law, as detailed above. One might wonder if it is possible to select a different kind of dispersion relation which can have more tangible effects in the infrared (IR). The answer to this question is affirmative. Indeed, let us consider a new dispersion relation of the type
\be\label{omnew}
\om^2 = k^2 \left(1+b_n\frac{k^n}{M^n}\right) f(k)+ \beta M^2 g(k)\ln\left(1+\frac{k^2}{\beta M^2}\right),
\ee
where $b_n$ and $\b$ are constants and $f(k)$ and $g(k)$ are dimensionless functions of the ratio $k/M$, with the following behavior. In the IR ($k/M\rightarrow 0^+$), we impose the logarithmic term to drop away, $f(k)\rightarrow 1$ and $g(k)\to 0$. In the UV ($k/M\rightarrow +\infty$), we impose the asymptotic form reproducing the entropy-area law, $f(k)\rightarrow 0$ and $g(k)\rightarrow 1$. For definiteness, we look at a specific example where $f(k)=1-g(k)=1-\tanh^2(k/M)$ [the case $f(k)=\exp(-k/M)$ would be somewhat more restrictive on the range of $n$]. Assuming $0<n<2$ and expanding in the IR, we get
\be\label{dire2}
\om^2 = k^2 \left(1+b_n\frac{k^n}{M^n} \right)+O\left(k^{n+3}\right).
\ee
The correction to the propagation speed coming from \Eq{dire2} is
\be\label{devbon}
\De v\simeq b_n\left(n+\frac{1}{2}\right)\left(\frac{\om}{M}\right)^n\,.
\ee
For $n=1$, one obtains an $O(k^3)$ correction to the standard dispersion relation, which leads to a linear correction on the propagation speed of the form \Eq{devqg}. The bound on $M$ is then \Eq{bouM}. However, Eq.\ \Eq{devbo} can be used to put bounds on $n$ much more stringent than the one derived above for the specific model studied in \cite{Pad98,Pad99} or for the generic $n=1$ case explored in \cite{ACEMN,EMNan}. Setting $b_n(n+1/2)=1$ and comparing Eqs.\ \Eq{devbo} and \Eq{devbon}, we find that $M> 10\,\text{TeV}$ if $n< 0.68$. Therefore, any dispersion relation with
\be\label{nbound}
0<n< 0.68
\ee
is allowed by the extant gravitational-wave observation and has a mass $M$ above the LHC scales. Remarkably, and contrary to Ref.\ \cite{EMNan}, we cannot compare any of these bounds with the much stronger ones coming from electromagnetic waves produced by distant objects such as gamma-ray bursts: the dispersion relations \Eq{dire} and \Eq{dire2} are typical only of the gravity sector.

To complement the lower bound \Eq{epin1}, we explore further consequences of quantum effects on black-hole physics. In Refs.\ \cite{Gid16,KoZh,CFP}, it was suggested to test quantum gravity by looking for new physics at the scales of the event horizon. Perhaps the most straightforward ``semiclassical'' feature which might affect the horizon structure is the backreaction from Hawking radiation. Such effect was first analyzed by Bardeen \cite{Bardeen:1981zz} and then thoroughly studied by York \cite{Yor83}. The upshot of these analyses is that the backreaction of quantum radiance produces a splitting between the event horizon and the apparent horizon which now lies {\it outside} the event horizon. The region within the two horizons is called the ``quantum ergosphere.'' The width $\De r$ of the quantum ergosphere is associated with a change in the observed irreducible mass
\be\label{rel1}
\De M_\text{BH}=\frac{\De r}{2}\,.
\ee
The irreducible mass undergoes the shift \cite{Arz05}
\be\label{rel2}
M_\text{BH}\to M_\text{BH}+\tilde\b\,\frac{M^2}{M_\text{BH}},
\ee
where $\tilde\b$ is a model-dependent parameter that can be computed explicitly. For instance, if we model the quantum ergosphere in terms of the quasinormal modes of the black hole \cite{Yor83}, the $l=2$ mode gives \cite{ABD}
\be\label{rel3}
\De r\simeq 3\times 10^{-5} \frac{M^2}{M_\text{BH}}\,.
\ee
Combining Eqs.\ \Eq{rel1} and \Eq{rel3} and identifying $\De M_\text{BH}$ with the correction in Eq.\ \Eq{rel2}, we obtain $\tilde\b\approx 1.5\times 10^{-5}$. However, in general, based on arguments relating the quantum ergosphere with log corrections to black-hole entropy \cite{Arz05}, we can allow $\tilde\b=O(1)$ values. 

To connect with observations, we impose the correction $\De M_\text{BH}=\tilde\b M^2/M_\text{BH}$ to be no greater than the experimental error $\de M_\text{BH}$ on the mass of the final black hole in the GW150914 merger:
\be
\De M_\text{BH}<\de M_\text{BH}\quad\Rightarrow\quad  M <\sqrt{\frac{M_\text{BH}\,\de M_\text{BH}}{\tilde\b}}\,.
\ee
For the LIGO merger, the mass was estimated to be $M_\text{BH}=62\pm 4\, M_\odot$ at the 90\% confidence level \cite{Abb16}. Since $1\,M_\odot\approx 10^{57}\,\text{GeV}$, for $\tilde \b\sim 10^{-5}-1$ we obtain a discouragingly high upper bound $M< 10^{58}-10^{60}\,\text{GeV}$. Clearly, this is of little use if we want to constrain quantum gravity efficiently.

Let us now consider the case of the quantum gravity-motivated ``phenomenological'' dispersion relation
\be\label{lqgdr}
\om \simeq k+b_2\frac{\om^3}{M^2}\,.
\ee
The second term in the right-hand side of Eq.\ \Eq{lqgdr} is the leading correction in the energy, as argued in Ref.~\cite{ACAP} based on an argument matching the logarithmic leading-order loop quantum gravity correction to the entropy area law. Note that a linear correction in $1/M$ to the dispersion relation would not reproduce the logarithmic modification to the Bekenstein--Hawking law, so that Eq.\ \Eq{lindi} is incompatible with quantum black-hole thermodynamics. On the other hand, Padmanabhan's \emph{Ansatz} \Eq{dire} or our variant \Eq{omnew} automatically generate this log correction. However, in contrast with scenarios of nonlocal field theories for quantum spacetime degrees of freedom of the type discussed above, this dispersion relation is expected to affect {\it universally} all matter and fields. After an expansion in the coefficient $b_2$, the group velocity reads
\be
v\simeq 1-3b_2\left(\frac{k}{M}\right)^2,
\ee
so that, replacing $k\simeq\om$, the net effect is
\be\label{devlqg}
\De v_\textsc{lqg} \simeq -3b_2\left(\frac{\om}{M}\right)^2.
\ee
This is of the same order of magnitude of the nonlocal case \Eq{devbh} and we get a constraint similar to \Eq{epin1}, unless $b_2\gg 1$ (which, however, would contradict our expansion).

We can also give a quick but stringent estimate on possible deformations of the loop-quantum-gravity dispersion relation of the form $\om \simeq \sqrt{1+\g}\, k$. The constant $\g$ is the leading correction in momenta to the standard dispersion relation. It has been formally calculated on a cosmological background from the expectation value of geometry operators \cite{AsDL} but we may assume that a similar correction (with different $\g$) would occur also for quantum states representing a Schwarzschild metric. In this case, it is easy to see that $\De v_\textsc{lqg} \simeq {\g}/{2}$, which immediately yields the bound $\g < 3\times 10^{-18}$.

To conclude, we have seen in several examples that the GW150914 gravitational-wave event cannot constrain efficiently the typical dispersion relations arising in quantum gravity. The IR correction $\om^2=k^2[1+O(k^n/M^n)]$ to the standard dispersion relation of gravitons is subject to the restriction \Eq{nbound} for realistic models where the mass $M$ is larger than the LHC energy scales but smaller than the Planck mass. Both this IR correction and the full dispersion relation \Eq{omnew} are completely \emph{ad hoc}, apart from requiring the recovery of the entropy-area law for black holes [which is failed by Eq.\ \Eq{lindi}]. Therefore, we do not regard Eqs.\ \Eq{lindi} and \Eq{omnew} as viable ways to see robust, rigorous quantum-gravity effects in the near future. 

On the other hand, the range \Eq{nbound} is compatible with the class of dispersion relations found in a model completely independent of quantum-gravity scenarios, the theory of multifractional spacetimes with $q$-derivatives \cite{qGW}. In that case, $n=1-\a$ is a parameter which depends on the fractional exponent $\a$ appearing in the measure describing an anomalous spacetime geometry with multifractal properties. This exponent lies in the range $0<\a<1$, the central value $\a=1/2$ being somewhat typical. For that theory, the GW150914 event does provide valuable information and useful bounds on the scale $M$ at which fractal effects become important. The results of \cite{qGW} contribute to illustrate the ultimate reason of the weakness of the constraints found in the present paper, which is neither the flatness of the background near the observer nor the very low frequency of the gravitational-wave signal. Rather, the actual reason is the form of the dispersion relation. The agent responsible for corrections to dispersion relations is the very texture of spacetime rather than local curvature effects. Thus, gravitational waves become a most direct testing ground for widely different models of exotic geometry. Future observations will hopefully tell us more about this story.

\paragraph*{Note added.} After the submission of this paper, we became aware of another work placing constraints on the quantum-gravity mass scale appearing in a modified dispersion relation for the graviton \cite{YYP}. Their dispersion relation (25) corresponds to our Eq.\ \Eq{dire2} with $\a_{\rm YYP}=2+n$ and $\mathcal{A}=M^{-n}$. Their Fisher analysis is based on frequencies $f=\om/(2\pi)=100\,{\rm Hz}$, corresponding to $\om\approx 630\,{\rm Hz}\approx 4.1\times 10^{-13}\,\text{eV}$ and $|\De v|<4.2\times 10^{-20}$. Then, the bounds \Eq{bouM}, \Eq{epin1}, and \Eq{nbound} are slightly improved to
\ba
&& M> 10^7\,{\rm eV}=10\,{\rm MeV}\,,\qquad n=1\,,\\
&& M> 6\times 10^{-3}\,{\rm eV}\,,\qquad n=2\,,\\
&& 0<n< 0.76\,.
\ea
These numbers agree, whenever a comparison is possible, with the findings of Ref.\ \cite{YYP} (see their Fig.\ 2). Note that these constraints do not change our main conclusion, also because the coefficients in our dispersion relations are phenomenological and may vary up to one order of magnitude, thus compensating a change in the frequency.

\section*{Acknowledgments.} We would like to thank G.\ Amelino-Camelia for insightful discussions, G.\ Nardelli for suggesting Eq.\ \Eq{omnew}, and the authors of Ref.\ \cite{YYP} for drawing our attention to their work and to the different choice of frequency. The work of G.C.\ is under a Ram\'on y Cajal contract.

\end{document}